\documentclass[prl,twocolumn,showpacs,aps]{revtex4-1}  
\usepackage{graphicx}  
\usepackage{dcolumn}   
\usepackage{bm}        
\usepackage{amsmath}   
\usepackage{amssymb}   
\usepackage{color}     

\newcommand{\comm}[2]{\ensuremath{\left[ #1,#2 \right] }}
\newcommand{\ket}[1]{\ensuremath{\left\vert #1 \right\rangle}}
\newcommand{\bra}[1]{\ensuremath{\left\langle #1 \right\vert}}

\newcommand{\abs}[1]{\ensuremath\left\vert #1 \right\vert}
\newcommand{\junits}{\ensuremath{\abs{\Delta}^{7/4}}}

\begin{document}
\title{Quantum quenches in two spatial dimensions using chain array matrix product states}
\author{A. J. A. James}
\affiliation{London Centre for Nanotechnology, University College London, Gordon Street, London WC1H 0AH, United Kingdom}
\email{andrew.james@ucl.ac.uk}
\author{R. M. Konik}
\affiliation{CMPMS Department, Brookhaven National Laboratory, Upton, New York 11973, USA}
\date{\today}
\begin{abstract}
We describe a method for simulating the real time evolution of extended quantum systems in two dimensions.
The method combines the benefits of integrability and matrix product states in one dimension to avoid several issues that hinder other applications of tensor based methods in 2D.
In particular it can be extended to infinitely long cylinders.
As an example application we present results for quantum quenches in the 2D quantum (2+1 dimensional) Ising model. 
In quenches that cross a phase boundary we find that the return probability shows non-analyticities in time.
\end{abstract}
\maketitle
The advent of ultra cold atomic gas experiments has led to a surge of interest in the time evolution and out-of-equilibrium behaviour of many-body quantum systems.
Much effort has been focused on one dimensional (1D) problems because these can be tackled by analytically tractable or highly accurate numerical methods.
Key questions that these studies have sought to elucidate are whether and how such systems thermalise after a sudden change, or `quantum quench' of a system's Hamiltonian; with particular emphasis on the role played by conserved charges in 1D integrable systems \cite{calabrese2006time,calabrese2007quantum,rigol2007relaxation,cassidy2011generalized,calabrese2011quantum,calabrese2012quantumI,calabrese2012quantumII,essler2012dynamical,caux2013time,fagotti2013finite,trotzky2012probing}.

Experiments however, are not limited to 1D and it is interesting to explore similar questions in two dimensions (2D) and above \cite{greiner2002collapse}.
Unfortunately there is no analogue in 2D of the aforementioned analytically exact 1D methods.
Numerical approaches using matrix product state (MPS) representations, so successful in 1D, suffer in 2D due to the `area law' growth of entanglement \cite{eisert2010colloquium,schollwock2011density}.
This growth reduces the efficiency of MPS (and related `tensor') algorithms and limits them to smaller system sizes.

Nonetheless MPS algorithms can be applied in 2D, by labeling lattice sites (usually in a zigzag fashion) to map to a 1D system \cite{stoudenmire2012studying}.
The cost is that nearest neighbor interactions in 2D are mapped to increasingly long ranged 1D interactions, imposing an increasing numerical burden.
Recently progress has been made in performing real time evolution on MPS with such long ranged Hamiltonians by two different routes \cite{zaletel2014time,haegeman2014unifying}.
Algorithms based on generalizations of MPS to higher dimensions, such as projected entangled pair states (PEPS) \cite{verstraete2004renormalization,murg2007variational}, make use of imaginary time evolution to find ground states \cite{phien2014faster}.
However these higher dimensional tensor methods have not been applied to real time evolution.

In this letter we demonstrate that \emph{real} time evolution is possible for large 2D systems by combining information
coming from exactly solvable models with a highly anisotropic MPS formulation.
Such an approach retains the contraction efficiency of matrix product states over other tensor methods, while avoiding the build up of long ranged interactions.
Our setup will be similar to that used in the density matrix renormalisation group (DMRG) studies described in Refs. \cite{konik2009renormalization,james2013understanding} except that here we are explicit in our use of MPS.
This change allows for straightforward implementation of algorithms other than DMRG, including those for time evolution and for accurately working with the thermodynamic limit.
In particular using time evolving block decimation (TEBD) \cite{vidal2004efficient} we demonstrate that we can study the time evolution after a quench of infinitely long cylinders, with sufficient circumference that we approach the 2D thermodynamic limit.
This includes strong quenches where we cross phase boundaries of a 2D quantum system.
\begin{figure}
\includegraphics[width=3.4in]{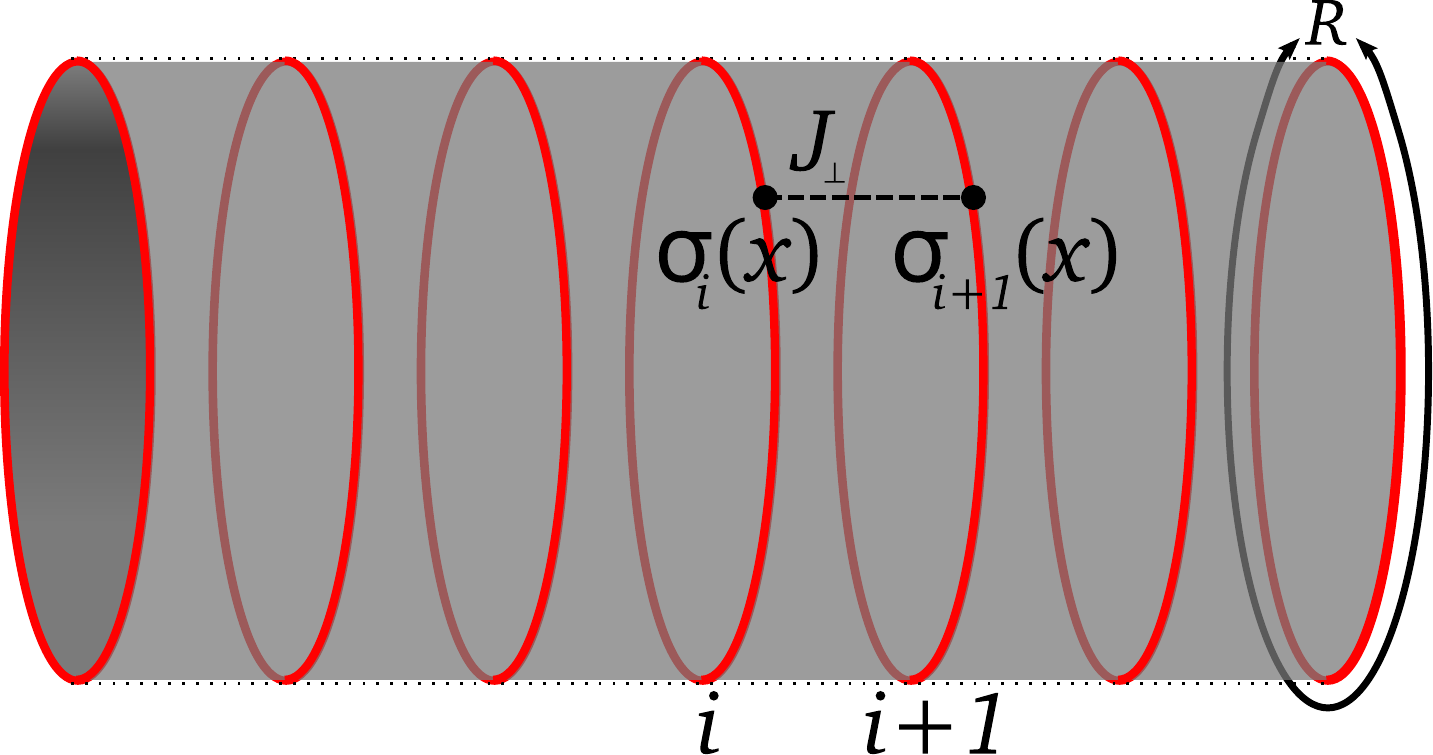}
\caption{Anisotropic setup for a 2D system as an array of $N$ chains of length $R$, coupled by an interaction $J_\perp$. The cylinder can be joined together at its ends to study toroidal systems.}
\label{fig:cylinder}
\end{figure}

\noindent {\textbf{Method:}}
At the core of our method is the wish to maximise the analytically exact input going into our MPS algorithm, while simultaneously controlling the growth of entanglement entropy.
The construction we use is depicted in Fig. \ref{fig:cylinder}: a coupled array of {\it exactly solvable} 1D subunits.
For each subunit, we have {\it exact} knowledge of the spectrum and matrix elements.
This exact knowledge means that we begin with the numerics already having accounted for much of the strong correlations of the system.
We emphasize our use of exactly solvable models as a building block is not much of a limitation to the method.  Such models are
ubiquitous in 1D, including Heisenberg spin chains, Luttinger liquids, and Hubbard models to name but a few \cite{essler2004applications,tsvelik1995quantum}.
In this framework, a state of a system of $N$ chains is written in MPS form via
\begin{align}
\ket{\Psi}_\mathrm{2D}=\sum_{\boldsymbol{\sigma}} A^{\sigma_1 [1]} \cdots A^{\sigma_N [N]}|\sigma_1\cdots\sigma_N\rangle
\label{eq:mps}
\end{align}
where each matrix $A^{\sigma_i [i]}$ is labelled by a chain $i$ and an eigenstate of that individual chain $\sigma_i$.
Like the single sites used in 1D MPS algorithms, we are able to manipulate these chain eigenstates 
because we know their energies and matrix elements for any relevant operator.

For ground and low-lying states of the system the entanglement entropy $S_E$ scales as the boundary `area', that is to say the chain length.
By keeping the chain length finite we can throttle the growth of $S_E$.  By partnering this with the fact that for the systems that we
will study, finite size effects are {\it exponentially suppressed}, we are able to keep $S_E$ small while remaining in the 2D thermodynamic
limit.  We have previously demonstrated the effectiveness of this methodology in equilibrium by studying a 2D quantum (i.e. $2+1$ dimensional) critical point \cite{konik2009renormalization,james2013understanding}.

The continuum 1D subunits will necessarily have an infinitely large Hilbert space. However if the system size $R$ is finite the spectrum is discrete, and we may truncate at a cutoff energy $E_c$.
This step is justified by appeal to the truncated conformal spectrum approach \cite{yurov1991truncated} where it has been observed over
a wide body of examples \cite{konik2011,lassig1991,lences2014} that for \emph{relevant} (in the renormalisation group sense) interchain
interactions, the low energy sector of a perturbed integrable system is formed primarily from (possibly strong) admixtures of low lying states of the unperturbed system.
Here we will focus on exactly such interchain perturbations.

Eq. \ref{eq:mps} differs from a MPS for a 1D system only in that the `physical indices' $\sigma$ may be large (see Table \ref{tab:num_chainstates} in \cite{suppmat}), requiring strict use of sparse matrices to maximise computational resources.
It is also important to take advantage of good quantum numbers and to perform matrix operations (e.g. singular value decompositions) in a block diagonal manner, to help preserve the sparse nature of the matrices and increase numerical efficiency.

MPS time evolution algorithms may then be implemented just as for a 1D system, including TEBD \cite{vidal2004efficient} and its infinite counterpart (iTEBD) \cite{vidal2007classical,orus2008infinite}.
For the former we may work with a torus or open cylinder geometry; the latter corresponds to an infinitely long cylinder.
Both algorithms decompose the time evolution operator $\exp[-i H t]$ into a product of $N_t$ time step operators, $t=N_t \tau$.
Each step is itself approximately decomposed into a product of two site (or chain) operations. 
The error at each step is proportional to the time increment $\tau$ raised to a power given by the order of the decomposition.

A more important source of error is the compression of the MPS after each step via Schmidt decompositions.
We compress by fixing a minimum singular value size, $s_{min}$: singular values smaller than this threshold value are discarded.
In this sense our algorithm is adaptive, as $\chi$, and the degree of encoded entanglement can grow.
`Lieb-Robinson' type arguments limit the rate of growth of $S_E$ after a quench \cite{calabrese2005evolution,bravyi2006,eisert2006}, but $\chi$ may grow exponentially, limiting the maximum timescales that can be reached.

For our 2D algorithm, forming the time evolution operator requires the exponentiation of a two chain Hamiltonian, which in turn necessitates the diagonalisation of the same object.
This is a numerically costly step, but need only be done once at the beginning, and the result stored for later use.

In this letter we present results for quenches in the 2D quantum Ising model:
\begin{align}
H_\mathrm{2DQI}=\sum_i \left[ H_{\mathrm{1D},i} + J_\perp \int_0^R \!\!\! \mathrm{d}x \ \sigma^z_i(x) \sigma^z_{i+1}(x) \right].
\end{align}
We represent the model as 1D Ising chains (of index $i$ and length $R$) coupled together with a longitudinal spin-spin 
interaction.
We take each chain $H_{\mathrm{1D},i}$ to be the continuum limit of the 1D lattice quantum Ising model---or transverse field Ising model (TFIM)---with Hamiltonian, 
$-J_\parallel\sum_l[\sigma^z_{i,l}\sigma^z_{i,l+1} + (1+g)\sigma^x_{i,l})]$ with $l$ an index along the chain.
In the continuum limit this reduces to a theory of a 1D Majorana field with mass $\Delta = gJ_\parallel$.
Analytic expressions for the spectrum of this theory 
and the spin matrix elements are detailed in Ref. \cite{fonseca2003ising}; we summarize the salient features in \cite{suppmat}.
Expanding the Majorana field in terms of fermionic modes $\psi^\dagger_{k_i}$ and $\psi_{k_i}$ (the continuum versions of the usual Jordan-Wigner lattice fermions) yields a quadratic chain Hamiltonian $H_{\mathrm{1D},i}=\sum_{k_i} \epsilon_{k_i} \psi^\dagger_{k_i} \psi_{k_i}$, with dispersion $\epsilon_{k_i}=\sqrt{\Delta^2 + k_i^2}$. We work in units such that the \emph{intrachain} velocity, $v$, is dimensionless and equal to unity. We also define a dimensionless interchain coupling $j_\perp=J_\perp \abs{\Delta}^{-7/4}$.
For disordered ($\Delta < 0$) chains a finite value of the interchain coupling $j_\perp$ leads to a 2D quantum (d=2+1) order-disorder transition at a critical value $j_\perp=j_c =0.185$ \cite{james2013understanding}.


 We compute the evolution of the postquench state using iTEBD and TEBD, with first and second order Trotter decompositions of the time evolution operator, and time steps $\tau$.
The error associated with such decompositions is dependent on $j_\perp$ and $\tau$, but even for the strongest quenches presented in this work we can choose $\tau$ small enough for convergence (see the supplementary material \cite{suppmat}).
For each set of parameters, we first establish that the numerical results are converged in $s_{min}$ or $\chi$ before increasing the cutoff $E_c$.
Convergence of the method in $s_{min}$ is demonstrated in \cite{suppmat}.
We have also checked the algorithm for two analytically tractable cases: the perturbative limit ($j_\perp \ll 1$) and a model of free fermionic chains with interchain hopping.
In both cases we find excellent agreement with our numerical results \cite{suppmat}.


\noindent {\textbf{Results:}}
\begin{figure}
\includegraphics[width=3.4in]{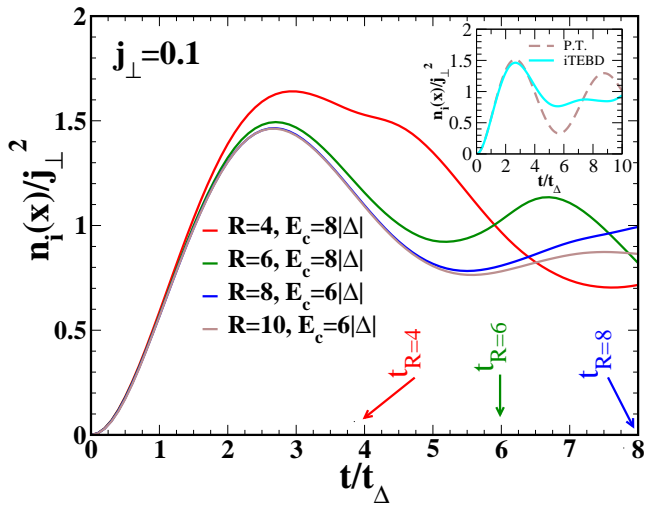}
\caption{Fermion occupation number, $n_i(x)$ scaled by interchain coupling, $j_\perp^2$.  We indicate
the time scale $t_R$ at which we expect the system postquench to see the effects of the finite circumference of the system.
Inset: $R=10$ iTEBD data compared with the perturbative result (P.T.) (dashed line).}
\label{fig:n_i_x_J0_1}
\end{figure}
\begin{figure}
\includegraphics[width=3.4in]{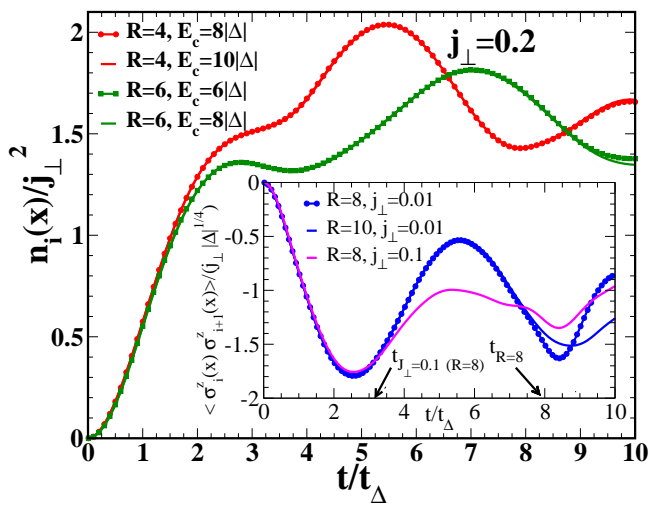}
\caption{Fermion occupation number, $n_i(x)$, scaled by interchain coupling, $j_\perp =0.2$, squared.
Curves for different $E_c$ are shown, corresponding to more than doubling the number of retained states in the chain spectrum. The agreement is excellent until the latest times, even though this quench crosses a critical point.
Inset: the nearest neighbor spin-spin correlation function showing scaling with $j_\perp$ and $R$.}
\label{fig:n_i_x_J0_2}
\end{figure}
In the following we present results of quantum quenches where the initial state of system corresponds to the $j_\perp=0$ ground state,
whereupon at $t=0$ we turn on a finite interchain coupling $j_\perp$.
We focus mainly on results for infinitely long cylinders, leaving a discussion of the effect of finite chain number, $N$, until the end.
We first address the question of what time scales we expect to feature in the quench. 
To provide a partial answer we turn to the quasiparticle causality picture of Refs. \cite{calabrese2005evolution,calabrese2006time,calabrese2007quantum}.
The energy imparted by the quench produces quasiparticle excitations which are entangled on a length scale $\abs{\Delta}^{-1}$ along the chain.
Intrachain scattering then only has an effect after a time, $t_\Delta=(2 v \abs{\Delta})^{-1}$.
On the other hand, the time scale governing interchain scattering can be estimated using Fermi's golden rule to be $t_{J_\perp}=\abs{\Delta}^{1/2}(J_\perp R)^{-2}$.
The final time scale of import is that encoding the chain length, $R$.
This scale, given by $t_R \sim R/2v= \abs{\Delta} R t_\Delta$, describes the time for two quasiparticles, created at the same point and moving in opposite directions, to travel around a chain and then meet again.
Hence there is a region, $t_\Delta,t_{J_\perp} < t < t_R$, where we may expect the time evolution to be representative of the 2D thermodynamic limit.
But for $t>t_R$ the finite nature of the chains' circumferences will play a role.
We stress that $t_R$ does not govern the time scale for revivals in the system.
Instead these occur on a much longer time scale, $t_{revival}\sim Nt_{J_\perp}$ where $N$ is the number of chains in the system.
Thus in our iTEBD simulations, we never expect to see strict revivals.

To illustrate these time scales in operation, we consider the occupation number, $n_{i}(x) =\psi^\dagger_{i}(x) \psi_{i}(x) $, 
for a fermionic mode on chain $i$, a simple measure of how the system departs from the initial state, for which $n_i(x)=0$.
In Fig. 2 we present how $n_i(x)$ evolves with time for a quench to $j_\perp = 0.1$.  On the basis of 
our perturbative results for very small $j_\perp$ \cite{suppmat}, we plot $n_{}(x)$ in units of $j_\perp^2$ for all four quenches presented.
These four quenches correspond to four different chain lengths, $R$. 

We see that at short times, the results for $n_i(x)/j_\perp^2$ collapse onto a single curve as a function of $ t/t_\Delta$.
As time increases, the curves cease to track one another.   The first to do this is the $R=4$ curve, then the $R=6$ curve, and then
finally the $R=8$ curve.   The time at which this happens corresponds, roughly, to $t_R$, the scale on which the quench
explores the finite length of the chain.
We expect a small departure from this time scale because a finite $j_\perp$ will renormalize the quasiparticle velocity $v=1$ in $t_R$.
We also see from the inset of Fig. 2 that the evolution at longer times is no longer described by perturbation theory.

In Fig. 3 we explore a quench to a $j_\perp$ which exceeds $j_c$, the critical coupling for the $2+1$ dimensional system.
Such a quench is among the most challenging numerically as the population of higher energy chain states becomes significant.
Concomitantly, the time evolution is most dependent on $E_c$ in this case.
Ramped, rather than sudden, quenches can be implemented with some possible advantages in this regard \cite{ramped}, though we have not yet explored this possibility.
Nonetheless in Fig. 3 we see that for a given chain length, $R$, we can find cutoffs, $E_c$ such that the time evolution is converged. 

It is also possible to calculate postquench correlations between the chains.
We show the nearest neighbor spin--spin correlation function as a function of time, $\langle \sigma^z_i(x,t) \sigma^z_{i+1}(x,t) \rangle$, for a selection of $R$ and $j_\perp$ in the inset of Fig. \ref{fig:n_i_x_J0_2}.
Our choice of $j_\perp>0$ favors antiferromagnetic correlations, producing the overall negative sign.
An expansion in small $t$ shows that this quantity is proportional to $j_\perp t^2$ allowing us to collapse the results onto a single curve at short times.
Here we see signatures of both the $t_{J_\perp}$ and $t_R$ scales.
In the inset we have marked the intrachain scattering time $t_{J_\perp}$, for the system with $R=8$ and $j_\perp=0.1$.
It is visible as the time that the $j_\perp=0.1$ and $j_\perp =0.01$ data begin to diverge.
We also mark the time scale $t_R$ at which the data for chains with $R=8, j_\perp=0.01$ begins
to diverge from that of $R=10, j_\perp=0.01$.

\begin{figure}
\includegraphics[width=3.4in]{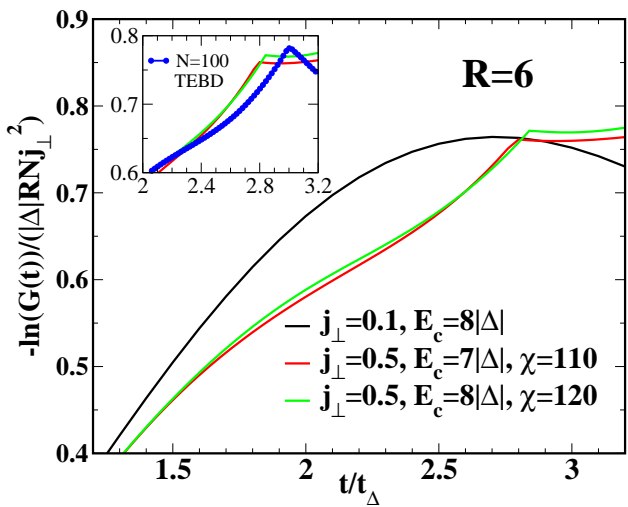}
\caption{Logarithm of return probability $G(t)$, for $R=6$ for $j_\perp=0.1,0.5$. Non-analytic behaviour is seen at short times for a quench to $j_\perp=0.5$. We find no non-analytic points for the corresponding quench to $j_\perp=0.1$, even at longer times up to $t=t_\Delta=10.0$ (not shown). Inset: comparison of the infinite chain number system data with a system with $N=100$ chains computed using TEBD. The first non-analytic point for the infinite cylinder forms the edge of a plateau, whereas for a finite number of chains it takes the form of a peak.}
\label{fig:loschmidt}
\end{figure}
To show that our method can handle non-trivial aspects of quenching through the critical coupling of the coupled chain system, we search for non-analyticities in the Loschmidt echo as a function in time.
The `Loschmidt echo' or overlap probability at a particular $t$ is the modulus squared of the overlap between the initial and time evolved state:
\begin{align}
G(t)=\left\vert \bra{\Psi_0} e^{-iH_\mathrm{2DQI}t} \ket{\Psi_0} \right\vert^2
\end{align}
where $\Psi_0$ is the ground state of the uncoupled chain system.
In 1D it is useful to define a per site rate function, $\ell(t)$ via $G(t)=\exp[-N\ell(t)]$.
Non-analyticities in $\ell(t)$ have been interpreted as `dynamical phase transitions', following an exact calculation of this quantity for the 1D TFIM \cite{heyl2013dynamical,pollmann2010dynamics,karrasch2013dynamical}.
The general association of such non-analytic points with equilibrium critical phenomena is contested \cite{fagotti2013dynamical,andraschko2014dynamical}, but we demonstrate analytically in low order perturbation theory\cite{suppmat} that for quenches to $j_\perp>0.27$ we expect non-analyticities in $G(t)$.
While this estimate for the value of $j_\perp$ is larger than $j_c$ -- because of the low order to which we took the computation -- it does suggest that simple perturbation theory for the quantity $G(t)$ can be used to estimate the phase boundaries in some 2D quantum systems.

In Fig. \ref{fig:loschmidt} we plot $\log G(t)$ for a quench to $j_\perp=0.5$ -- a value of $j_\perp$ where we should see non-analyticities.
In 2D this quantity scales with system volume $RN$, as does its 1D counterpart \cite{heyl2013dynamical}.
It also scales with $j_\perp^2$.
As expected we find non-analytic behaviour for this quench, within the time window we are able to simulate, and see that the non-analyticity has the same qualitative structure for both $E_c=7\abs{\Delta}$ and $8\abs{\Delta}$.
For comparison we plot $\log G(t)$ for a quench to $j_\perp = 0.1$, where in contrast we find that this quantity is smooth within our simulation window.
We remark that non-analyticities appear for the same quantity with $j_\perp=0.2$ (not plotted), just above $j_c=0.185$, but they first occur only at the edge of the attainable times with iTEBD.

Finally we consider the case of finite length and open boundary conditions.
The TEBD algorithm is slower by approximately a factor of $N$ due to the loss of translational invariance along the cylinder.
We find negligible effect, for finite $N\gtrsim 10$ and $i$ away from the ends of the cylinder, on the results for local quantities such as $n_i(x)$ (up to the time scales we reach).
However this is not true for the Loschmidt echo (a global measure), especially when $\vert j_\perp \vert >j_c$.
The inset of Fig. \ref{fig:loschmidt} shows the difference between the iTEBD and $N=100$ results for $R=6,j_\perp=0.5$.
While there is excellent agreement up to $t \sim t_\Delta$ (not shown), afterwards there is a clear change in the non-analytic point structure.
We also find that this effect is even more pronounced for very small $R$ and large $N$ (where our model reduces to a single 1D TFIM), suggesting that boundary conditions have a non-negligible effect on the Loschmidt echo even for large systems.
This last result has important consequences for possible experimental investigations.
\noindent {\textbf{Conclusions:}}
We have demonstrated a robust method to compute dynamical behaviour in 2D quantum (d=2+1) systems after a quench, which we intend to use to study other systems including coupled quantum wires (i.e. coupled Luttinger liquids) and Heisenberg chains.
The algorithm should prove especially useful when interpreting non-equilibrium cold atom \cite{JG,JG1} and pump-probe experiments in the cuprates \cite{DalConte30032012,Rameau1}.

We wish to acknowledge enlightening discussions with John Cardy, Fabian Essler, Andrew Goldsborough, Israel Klich, Anatoli Polkonikov, Rudolf R\"{o}mer and Steve Simons.
This work was supported by the Engineering and Physical Sciences Research Council (grant number EP/L010623/1) and the 
US Department of Energy, Office of Basic Energy Sciences under Contract No. DE-AC02-98CH10886.
\section{Supplemental Material}
\subsection{Free Fermions}
\begin{figure}
\includegraphics[width=3.4in]{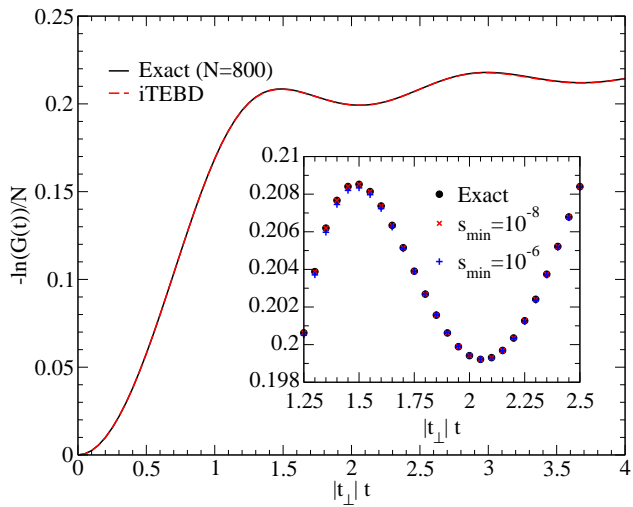}
\caption{Logarithm of return probability, $G(t)$, for the free fermion quench described in the text, with hopping parameter $t_\perp=0.5$. Both the exact result, using a determinant method with $N=800$ chains, and the iTEBD result are plotted. Inset: enlarged region showing the determinant calculation and iTEBD with different $s_{min}$ parameter.}
\label{fig:free}
\end{figure}

In this section we describe our method applied to an exactly solvable quantum model in 2D.
Consider a free Majorana field $\psi=\psi^\dagger$, $\{\psi(x,t),\psi(x',t) \}=\delta(x-x')$ with mass $\Delta$, confined to a ring and with a Lorentz invariant action
\begin{align}
S=\int dt \int_0^R dx \bar{\psi} (i \gamma^\mu \partial_\mu -\Delta)\psi,
\end{align}
where $\mu=0,1$ refers to time and space coordinates, $\bar{\psi}=\psi^\dagger \gamma^0$ and $\gamma^{0,1}$ are suitable 2D Dirac matrices. The field has an expansion in fermionic modes $\comm{a_n}{a_m^\dagger} = \delta_{m,n}$
\begin{align}
\psi=\sum_n &\sqrt{\frac{m}{2 \epsilon_n R}} e^{\theta_n/2} \big( \omega a_n e^{-i(t \epsilon_n -x p_n)}+ \nonumber \\
& \omega^\star a_n^\dagger e^{i(t \epsilon_n -x p_n)} \big), \nonumber \\
\bar{\psi}=-\sum_n &\sqrt{\frac{m}{2 \epsilon_n R}} e^{-\theta_n/2} \big( \omega^\star a_n e^{-i(t \epsilon_n -x p_n)}+  \nonumber \\
& \omega a_n^\dagger e^{i(t \epsilon_n -x p_n)}\big).
\end{align}
Here $\omega=e^{i \pi/4}$, and $\epsilon_n=\Delta \cosh \theta_n$, $p_n=\Delta \sinh \theta_n$. The momentum can take discrete values $p_n=2\pi n/R$ for integer $n$, and $\epsilon_n =\sqrt{\Delta^2 + p_n^2}$.
Note that in this case we are not mapping from a putative spin system using the Jordan-Wigner transformation, so there is no separation of the spectrum into Ramond and Neveu-Schwarz sectors.
Using the mode expansion we obtain the chain Hamiltonian
\begin{align}
H_{\text{1D},\ell}=\sum_n \epsilon_n a^\dagger_{n,\ell} a_{n,\ell},
\end{align}
with chain index $\ell$.
We can build a 2D quantum system from an array of $N$ of these chains coupled with a nearest neighbor interaction (here we assume our 2D system is a torus)
\begin{align}
H_\text{int}= -\sum_{\ell,n} t_{\perp} \frac{\Delta}{\epsilon_n} \big(a_{n,\ell}^\dagger a_{n,\ell+1} +\text{H.c.}\big)
\end{align}
where the coupling strength is $t_\perp$.
This chain array system $H_\text{free}=\sum_{\ell} H_{\text{1D},\ell} + H_\text{int}$ can be solved trivially by Fourier transformation from chain index $\ell$ to momentum $k_m=2\pi m/N$ (note this is transverse to the momentum index $n$ \emph{along} the chains).
\begin{align}
a^\dagger_\ell &= \frac{1}{\sqrt{N}}\sum_m e^{ik_m \ell} a^\dagger_\ell \\
\{a_{k_m}, a^\dagger_{k_{m'}}\}&= \delta_{k_m,k_{m'}}\\
H_\text{free}&=\sum_{n,m} \big(\epsilon_{n} -\frac{2\Delta t_\perp}{\epsilon_n} \cos k_m \big) a^\dagger_{n,m} a_{n,m}
\end{align}
The Hamiltonian is diagonal in $n$ and $m$: if so desired it can be treated as either a set of $N$ 1D uncoupled bands indexed by $m$, or infinitely many 1D bands, indexed by $n$.
In either case the system can then be treated by standard 1D MPS methods, because the lack of coupling between bands means that one needs to keep only those eigenstates from the spectrum of $H_{\text{1D},\ell}$ that are present in the initial state of the chain array.
Beyond this the cutoff $E_c$ does not play a role.
For example consider the initial state
\begin{align}
\ket{\Phi}=\prod_{i=0}^{\frac{N}{2}-1}  \frac{1}{\sqrt{2}} \big(\ket{0}_{2i} \ket{n=0}_{2i+1} +\ket{n=0}_{2i} \ket{0}_{i+1}\big),
\end{align}
in which alternating pairs of chains are entangled, with a superposition of ground states ($\ket{0}_i$) and lowest excited states ($\ket{n=0}_i$) on the chains.
Evolving this state under $H_\text{free}$ does not involve any other states from the chain spectrum.
For such an evolution the return probability can be calculated exactly using a determinant method, yielding
\begin{align}
G(t)&= \Big \vert\text{det} \big( M + (1-M) R \exp \{-i h t\} R \big)\Big \vert^2, \\
M&=\text{diag}(0,1,0,1,0,\cdots), \nonumber \\
R& =
\left(
\begin{array}{rrrrc}
  1& 1  &   & &\\
  1 & -1  & & &\\
  &   & 1  & 1& \\
  & & 1 & -1 & \\
  & & & & \ddots 
\end{array}
\right), \nonumber \\
h& =
\left(
\begin{array}{ccc}
  \Delta & -t_\perp  &   \\
 -t_\perp & \Delta  & \ddots  \\
  &  \ddots & \ddots   
\end{array}
\right),
\end{align}
with $N \times N$ matrices.
We compare this result (evaluated for a torus of $N=800$ chains) with the iTEBD result computed with our code implementing $H_\text{free}$ in Fig. \ref{fig:free}.
Note that in this case a change in $\Delta$ can be absorbed into a simultaneous rescaling of $t_\perp$ and $t$, so the hopping $t_\perp$ sets the time scale.
%
\subsection{Further details of method}
The structure of the spectrum of the continuum limit Ising chain is detailed in Ref. \cite{fonseca2003ising} but we give a brief description here for convenience.
The spectrum splits into two sectors, Neveu-Schwarz (NS) and Ramond (RM).
The energy of a state with a particular fermion configuration is given by 
\begin{align}
E(\{n_s\})=E_s+\sum_{\{n_s\}} E_{n_s} \\
E_{n_s}=\sqrt{\Delta^2+\Big(\frac{2\pi n_s}{R}\Big)^2}
\end{align}
where $s=\mathrm{NV}$ or $\mathrm{RM}$ and $E_s$ is the vacuum energy (different in the two sectors).
For the disordered phase of a chain, $\Delta<0$, states with even numbers of particles (including the 0 particle vacuum state) are in the NS sector ($n_s \in \mathbb{Z}$), while odd particle states are in the Ramond sector ($n_s \in \mathbb{Z} /2$)
The spin operator, $\sigma^z$, is off diagonal in sector, so that on an individual chain RM states are only scattered into NS states and vice-versa by the $J_\perp$ term.
This fact that makes perturbative calculations significantly easier.
As a consequence the overall sector (whether there is a odd or even number of Ramond chains) is conserved by $H_{\mathrm{2DQI}}$.
The total sector and momentum (along the chain direction) for two chains is also conserved by the two chain time evolution operator.
These conservation laws are useful for performing matrix operations by sub blocks.

The fermionic representation is symmetric with respect to the spin direction, as is our pre-quench state, and $H_{\mathrm{2DQI}}$ itself does not break this symmetry.
Hence the local magnetization $\langle\sigma^z_i(x)\rangle$ is always zero.

We implement iTEBD \cite{vidal2007classical} using the alteration due to Hastings \cite{hastings2009light} that improves numerical robustness by removing the need to divide by very small singular values.
Before computing expectation values the iTEBD transfer matrix must be `orthogonalised' as described in Refs. \cite{orus2008infinite,mcculloch2008infinite,schollwock2011density}.

\begin{figure}
\includegraphics[width=3.4in]{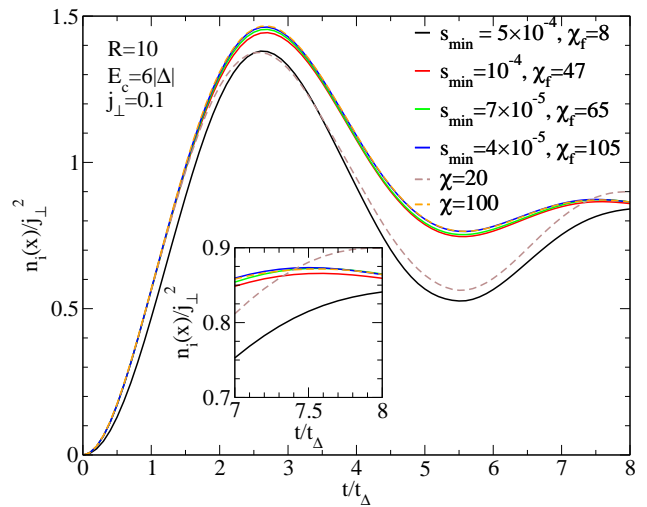}
\caption{Convergence in $s_min$ for $R=10$, $J_\perp =0.1$, using a smallest retained singular value criterion, $s_min$ (the largest bond dimension used during the calculation is given as $\chi_f$. Also shown for comparison is data collected using a fixed bond dimension, $\chi$.}
\label{fig:conv}
\end{figure}

Instead of imposing a fixed bond dimension, we perform TEBD and iTEBD using a cutoff on the minimum singular value that is retained, $s_{min}$. 
This translates to a minimum eigenvalue of the reduced density matrix, $\rho_{min}=s_{min}^2$.
The advantage over fixed bond dimension is that when working with large matrices, one may drop unwanted singular values as individual sub blocks are processed, as opposed to recording the full results of the singular value decomposition and sorting by magnitude before truncating.
In principle the maximum value of $S_E$ that can be captured in this way is $-2 \log s_{min}$.
In Fig. \ref{fig:conv} we give an example of convergence in this parameter and compare with a version of the algorithm that uses a fixed bond dimension throughout.

\begin{figure}
\includegraphics[width=3.4in]{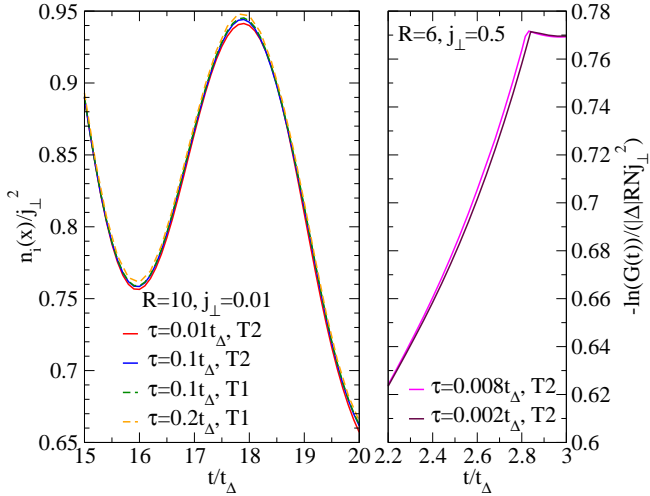}
\caption{Convergence in the time step $\tau$ (in units of $t_\Delta$). Left panel: In the perturbative ($j_\perp \ll 1$) limit results converge extremely quickly as a function of $\tau$, for both 1st order ($T1$) and 2nd order ($T2$) Trotter decompositions. Right panel: At the other extreme, for $j_\perp>j_c$ we see that the sensitive placement of non-analytic points is converged for sufficiently small $\tau$.}
\label{fig:timestep}
\end{figure}

We mainly use the second order Trotter decomposition in this work, and find it sufficient for the time scales we are able to reach before bond dimension, $\chi$, becomes the limiting factor.
For very weak quenches however, a first order Trotter decomposition still gives good results, even to quite large times.
This is because the leading error term in the decomposition is proportional to $J_\perp$.
A more sensitive test is whether the sharp non-analytic features in the return probability (Loschmidt echo) for $j_\perp=0.5$ are well converged in the time step $\tau$.
Figure \ref{fig:timestep} shows the logarithm of the return probability for $\tau=0.008 t_\Delta$ and $0.002t_\Delta$ using second order Trotter decompositions.
There is negligible difference except in the immediate approach to the non-analytic point, and even in this case the difference is less than $1\%$. 

\subsubsection{Discussion of the effects of the cutoff}
\begin{table}[ht]
\caption{The number of chain states kept (including the ground state) for various combinations of $R$ and $E_c$ ($\Delta=-1$). }
\begin{center}
\begin{tabular}{c|ccccc}
&\multicolumn{5}{c}{$E_c/\abs{\Delta}$}\\
$R$ &4 &5 &6 &8 & 10 \\
\hline
2  & 5  & 5  & 5  & 12  & 19 \\
4  & 11 & 14 & 19  & 43  & 90 \\
6  & 19 & 33 & 52  & 124 & \\
8  & 29 & 55 & 103 & 308 & \\
10 & 44 & 97 & 172 &     & \\
12 & 63 &     &     & 
\end{tabular}
\end{center}
\label{tab:num_chainstates}
\end{table}%
The number of chain states increases approximately exponentially with $E_c$ and $R$ for $E_c > \Delta$. In Table \ref{tab:num_chainstates} we show the number of chain states kept for a range of chain lengths and cutoffs.
It is clear that changing $E_c$ from $4\abs{\Delta}$ to $8\abs{\Delta}$ (for example) has a much more dramatic effect on the number of included states at $R=8$ than at $R=4$. However with quenches below the critical coupling $J_c$, we find that in general we see very little difference between $E_c=6\abs{\Delta}$ and $E_c=8\abs{\Delta}$ for $R\ge4$.
\subsubsection{Small R limit}
For a single chain there should be a crossover to effective 0D (0+1 dimensional) behaviour when the correlation length is of order the system size, $\vert \Delta \vert^{-1} \sim R$.
When $\vert \Delta\vert R \ll 1$ only the chain ground state and lowest excited state survive with the energies of higher excited states scaling as multiples of $(\vert\Delta\vert R)^{-1}$.
For $\Delta<0$ the ground and first excited states are the Neveu-Schwarz vacuum and the zero momentum single particle Ramond states respectively.
In this case the Hamiltonian of the coupled chain system becomes
\begin{align}
\lim_{R\to 0} H_{\mathrm{2DQI}} = \sum_i
& \left[
\left(
\begin{array}{cc}
 \vert \Delta \vert +E_{\mathrm{RM}} &0    \\
 0 & E_{\mathrm{NS}}  
\end{array}
\right)_i
\right.  \nonumber \\
&\left. +J_\perp R
\left(
\begin{array}{cc}
0 & M    \\
 M & 0
\end{array}
\right)_i
\left(
\begin{array}{cc}
0 & M    \\
 M & 0
\end{array}
\right)_{i+1}
\right],
\end{align}
where $M= \bra{\mathrm{RM},k=0} \sigma^z \ket{\mathrm{NS,vac}} \in \mathbb{R}$.
This can be written as the Hamiltonian of a single 1D lattice quantum Ising chain (up to some unimportant constants):
\begin{align}
\lim_{R\to 0} H_{\mathrm{2DQI}} =H_{\mathrm{1DQI}}=\sum_i \big( \tilde{h} \tilde{\sigma}^z_i +\tilde{J} \tilde{\sigma}^x_i \tilde{\sigma}^x_{i+1} \big),
\end{align}
where $\tilde{\sigma}^x, \tilde{\sigma}^z$ are the usual Pauli matrices and we make the identifications
\begin{align}
\tilde{h} & = \frac{\vert \Delta \vert +E_{\mathrm{RM}}-E_{\mathrm{NS}}}{2},\\
\tilde{J} & = J_\perp R M^2.
\end{align}
This is a useful check of the code, as it is easy establish if one recovers the correct 1D behaviour, including the $1+1$ dimensional phase transition when $\tilde{h}=\tilde{J}$.
For example at $R=1$, and by using $E_c$ to restrict the number of chain states to two, we are able to successfully reproduce the predicted positions of non-analytic points, for quenches of the 1D quantum Ising model through its critical point \cite{heyl2013dynamical}.
\subsection{Perturbation Theory}
\begin{figure}
\includegraphics[width=3.4in]{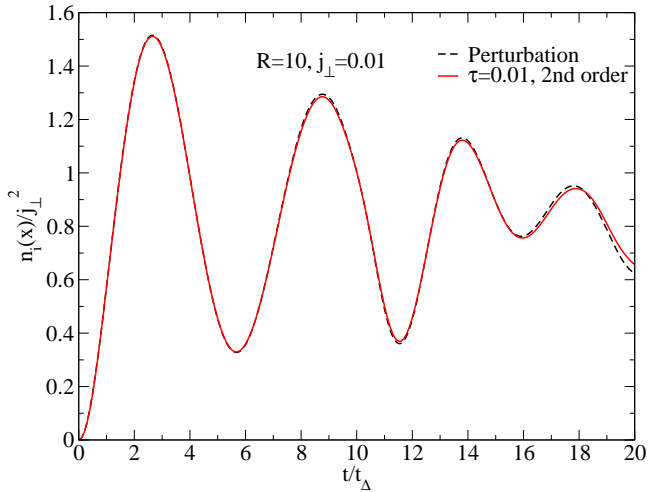}
\caption{Fermion occupation number, $n_i(x)$, at position $x$ on chain $i$, scaled by the interchain coupling, $J_\perp =0.01$, squared. Differences as a function of cutoff are negligible for $E_c>4\abs{\Delta}$.
Dashed line: the perturbative result for $R=10$.}
\label{fig:n_i_x_J0_01}
\end{figure}
In the limit of small interchain coupling $ \left\vert J_\perp/J_c \right\vert \ll 1$ a perturbative expansion is appropriate.
We use unitary perturbation theory, following Ref. \cite{kollar2011generalized}, in order to avoid spurious secular terms that grow in time without bound.
The expectation of an operator $A$ at time $t$ after the quench (assuming that the pre quench Hamiltonian commutes with it, $\left[ H_{1\mathrm{D},i},A\right]=0$) is given to order $J_\perp^2$ by
\begin{align}
\langle A(t) \rangle & = \langle A(0) \rangle + 4 J_\perp^2 \int_{-\infty}^{\infty} \!\!\! \mathrm{d} \omega F(\omega) \frac{\sin^2 (\omega t /2)}{\omega^2}, \\
 F(\omega)&=\sum_\Phi \Big \vert \bra{\Psi_0} \sum_i \sigma^z_i \sigma^z_{i+1} \ket{\Phi} \Big \vert^2 \\
 & \qquad \times \bra{\Phi} A \ket{\Phi} \delta\big(\omega - (E_\Phi-E_{\Psi_0})\big),
\end{align}
where $\ket{\Phi}$ is a tensor product of unperturbed chain states, $\ket{\phi_1} \otimes \ket{\phi_2} \otimes \cdots \otimes \ket{\phi_N}$.
In this work we choose the pre quench state to be the ground state of the uncoupled chain system,
\begin{align}
\ket{\Psi_0}=\prod_{\otimes_i} \ket{0_i},
\end{align}
where $\ket{0_i}$ is the ground state of chain $i$.
For the operator $A$ we consider $n_{i,k}$, the occupation number for a fermion on chain $i$ with momentum $k$ (along the chain).
With these choices the only states that contribute to $F(\omega)$ for the quantum Ising system are tensor products of chain vacuum states, with a single nearest neighbor pair of chain excited states:
\begin{align}
\ket{\Phi_+}&=\cdots \otimes \ket{0_{i-1}} \otimes \ket{\phi_i} \otimes \ket{\phi_{i+1}} \otimes \ket{0_{i+2}} \otimes \cdots, \nonumber \\
\ket{\Phi_-}& =\cdots \otimes \ket{0_{i-2}} \otimes \ket{\phi_{i-1}} \otimes \ket{\phi_{i}} \otimes \ket{0_{i+1}} \otimes \cdots.
\end{align}
Using these states we find 
\begin{align}
F(\omega)=&R^2 \Big [\sum_{\substack{\phi_{i+s},\phi_i \\ s=\pm1}}'  \Big \vert \bra{0_{i}} \sigma^z_i \ket{\phi_{i}} \bra{0_{i+s}} \sigma^z_{i+s} \ket{\phi_{i+s}} \Big \vert^2 \nonumber \\
& \times  \delta(\omega-(E_{\phi_{i}}+E_{\phi_{i+s}}-2E_0)) \bra{\phi_{i}} n_{i,k} \ket{\phi_{i}} \Big],
\end{align}
where $E_\phi$ is the energy of the (unperturbed) chain state $\ket{\phi}$. 
The restriction on the sum indicates that the momenta of the chain states $\ket{\phi_i}, \ket{\phi_{i\pm1}}$ must sum to zero.
For the expectation at time $t$ we obtain
\begin{align}
\langle n_{i,k} (t)\rangle&= (2 J_\perp R)^2 \!\!\!\! \sum_{\substack{\phi_{i+s},\phi_i \\ s=\pm1}}' \! \Big \vert \bra{0_{i}} \sigma^z_i \ket{\phi_{i}} \! \bra{0_{i+s}} \sigma^z_{i+s} \ket{\phi_{i+s}} \Big \vert^2 \nonumber \\
&\times\frac{\sin^2 (t E_{i+s,i}/2)}{E_{i+s,i}^2} \bra{\phi_{i}} n_{i,k} \ket{\phi_{i}}  \label{eq:perturbative}.
\end{align}
Here $E_{i+s,i}=E_{\phi_{i}}+E_{\phi_{i+s}}-2E_0$ and the restriction on the sum is as above.
The sum over $s$ will contribute a simple factor of $2$ unless the system is an open cylinder and $i=1,N$, in which case one of the sums vanishes due to the missing nearest neighbor.

We now make some remarks about the result at order $J_\perp^2$.
For all $i,k$ Eq. \ref{eq:perturbative} is a sum of oscillatory terms with no `decay' even in the thermodynamic limit.
There are no boundary effects, excepting the trivial factor of $2$ described above due to the different number of nearest neighbors.
With disordered chains, for which the chain ground state is in the Neveu-Schwarz sector, only excited chain states of the Ramond sector will contribute.
Consequently at this order $\langle n_{i,k} \rangle$ will be zero for half integer momenta, $k=2\pi (n+1/2)/R$.

Using the above we calculate the occupation numbers perturbatively for a large range of $k$ by evaluating the sums over states numerically.
From these results it is simple to to find the position space occupation:
\begin{align}
R n_i(x) = \int_0^R \!\! \mathrm{d}\tilde{x} \ n_i(\tilde{x})=\sum_k n_{i,k},
\end{align}
using translational invariance along the chains, provided the $n_{i,k}$ drop off sufficiently rapidly with $k$.
We show our results for $J_\perp=0.01$ together with the perturbative curve (dashed curve) for $R=10$ in Fig. \ref{fig:n_i_x_J0_01}.
The agreement is excellent at short times and still very good at longer times for $n_i(x)$.

\subsection{Estimate of Critical Coupling, $J_c$}
In this portion of the supplemental material we argue that the appearance of non-analyticities in the return
probability post-quench can be used to estimate the value of the critical coupling $J_c$ marking the phase transition (although
see discussion below).
To this end we employ the observation of \cite{heyl2013dynamical} that in quenching from an initial value of the coupling, $J_{\perp i}$,
to a final value of the coupling, $J_{\perp f}$, non-analyticities appear in the return probability whenever a fermionic mode 
(of the post-quench Hamiltonian) has an occupation of at least $1/2$, i.e. appears with an occupation corresponding
to either infinite or negative temperatures.  In the case considered in \cite{heyl2013dynamical}, these athermal occupations
only occurred if the coupling crossed a phase boundary.  The appearance of the non-analyticities can then be used to estimate the location of these
boundaries.  However it is at least possible that in general interacting models (such as
the XXZ spin chain considered in \cite{andraschko2014dynamical,fagotti2013dynamical}, such occupations can be induced without crossing
a phase boundary.  We will use low order perturbation theory to estimate the coupling $J_c$ at which athermal mode occupations appear,
noting that because of the work of \cite{andraschko2014dynamical,fagotti2013dynamical} that this $J_c$ may not correspond to the critical
coupling determining the equilibrium phase transition in the two dimensional quantum Ising model.

The modes that we will consider in this argument take the form 
\begin{equation}
\psi^\dagger_{k_x,k_y}(J_\perp)= \frac{1}{\sqrt{N}}\sum_j e^{ik_y j}A^\dagger_{j,k_x} + {\cal O}(J_\perp) + \cdots
\end{equation}
where $A^\dagger_{j,k_x}$ is an operator on the j-th chain that creates a fermion with momentum $k_x$ along the chain.
We then want to find the minimum value of $J_\perp$ such that
$$
\langle n_{k_x,k_y} \rangle = \langle i|\psi^\dagger_{k_x,k_y}(J_\perp)\psi_{k_x,k_y}(J_\perp)|i\rangle = 1/2,
$$
where $|i\rangle$ is the initial state of the quench (here the ground state of the system for $J_\perp = 0$).
The mode for which this will first occur is $(k_x,k_y)=(k_{x,min},0)=(2\pi/R,0)$ as this is the mode with the lowest energy that couples to the perturbation, and so is 
easiest to drive athermal.

To compute $\langle n_{k_{x,min},0}\rangle$ we expand it in terms of the eigenstates $\{|s\rangle\}$ of the post-quench Hamiltonian
\begin{equation}
\langle n_{k_{x,min},0}\rangle= \sum_s |\langle s|i\rangle|^2 \langle s|n_{k_{x,min},0}|s\rangle.
\end{equation}
We will suppose that this sum is dominated by states involving at most one fermion on any given chain.  The only
such state $|s\rangle$ that then contributes to this sum is
$|k_{x,min},0;-k_{x,min},0\rangle = \psi^\dagger_{k_{x,min},0}(J_\perp)\psi^\dagger_{-k_{x,min},0}(J_\perp)|0\rangle$.  While we cannot write down an 
exact expression for this state as a function of $J_\perp$, we are able to write down to second order the contribution to this
state coming
from the $J_\perp=0$ vacuum $|0\rangle$ -- the only part that matters in computing the overlap $\langle s|i\rangle$.  
To second order we have
\begin{eqnarray}
|k_{x,min},0;-k_{x,min},0\rangle &=& 
\psi^\dagger_{k_{x,min},0}(0)\psi^\dagger_{-k_{x,min},0}(0)|0\rangle \cr\cr
&& \hskip -1in + \frac{\bar\sigma^2 J_\perp}{E_{k_{x,min}}^2}(i-\frac{2J_\perp\bar\sigma^2}{E^2_{k_{x,min}}}+{\cal O}(J_\perp^2))|0\rangle
+ \cdots.
\end{eqnarray}
We use here the conventions and notation of \cite{fonseca2003ising}.  In particular $\bar\sigma=1.35783834\ldots$
and $E_{k_x}^2 = \Delta^2+k_x^2$.
Thus to second order in $J_\perp$ we have 
\begin{equation}
\langle n_{k_{x,min},0}\rangle= \frac{|\Delta|^{1/2}\bar\sigma^4J_\perp^2}{E_{k_{x,min}}^4} (1+\frac{4|\Delta|^{1/2}\bar\sigma^4J_\perp^2}{E_{k_{x,min}}^4}).
\end{equation}
For $R \to \infty$ the value of $j_\perp=J_\perp/\junits$ at which $\langle n_{k_{x,min},0}\rangle = 1/2$ is $j_\perp=0.27119\ldots$.  While this value of $j_\perp$ is considerably
larger than the value of $j_c = 0.185$ given in \cite{konik2009renormalization} for where the equilibrium phase transition
occurs, we can at least partially attribute this difference to the neglect both of terms higher order in $J_\perp$ in this computation 
as well as $J_\perp=0$ states involving more than one fermion per chain.

\begin{thebibliography}{41}
\bibitem [{ram()}]{ramped}%
  \BibitemOpen
  \href@noop {} {}\bibinfo {howpublished} {A sudden quench pumps more energy
  into the system relative to a ramped quench, and therefore will populate
  higher energy modes, reducing the efficiency of the algorithm.}\BibitemShut
  {Stop}%
\end{thebibliography}
%
\end{document}